\documentclass[letter]{aa}  
\usepackage{graphicx}
\usepackage{txfonts}

\usepackage{natbib}
\usepackage{amsmath} 
\usepackage{amssymb}
\usepackage{mathtools}
\usepackage{microtype}
\usepackage{lineno}

\definecolor{CiteRed}{RGB}{110, 0, 0}
\usepackage{multirow}
\usepackage{array}
\usepackage{soul} 
\usepackage{makecell}
\usepackage{siunitx}
\usepackage{color}      
\usepackage{orcidlink}  
\usepackage{graphicx}
\usepackage{subcaption}
\usepackage{lipsum}
\usepackage{verbatim}
\graphicspath{ {img/} {plot/} {fig/} }
\defcitealias{Guidorzi24}{G24}

\usepackage{hyperref}
\hypersetup{
    breaklinks,
    colorlinks,
    citecolor=CiteRed,  
    linktoc=page,
    pdftitle={Distribution of the number of peaks within a long gamma-ray burst:
the full \textit{Fermi}/GBM catalogue},
    pdfkeywords={GRBs},
    pdfauthor={Romain Maccary},
    pdfcreator={\LaTeX}
}

\begin{document} 

   \title{Distribution of the number of peaks within a long gamma-ray burst: The full \textit{Fermi}/GBM catalogue}

   \author{R.~Maccary~\inst{1,2}\fnmsep\thanks{\texttt{mccrnl[at]unife[dot]it}} \orcidlink{0000-0002-8799-2510}\and   
           M.~Maistrello~\inst{1,2}\orcidlink{0009-0000-4422-4151}\and
            C.~Guidorzi\inst{1,2,3}\orcidlink{0000-0001-6869-0835}\and
           M.~Sartori~\inst{1}\orcidlink{0009-0007-3573-5305}\and
           L.~Amati~\inst{2}\orcidlink{0000-0001-5355-7388}\and 
           L.~Bazzanini~\inst{1,2}\orcidlink{0000-0003-0727-0137}\and
           M.~Bulla~\inst{1,3,4}\orcidlink{0000-0002-8255-5127}\and
           A.~E.~Camisasca~\inst{7}\orcidlink{0000-0002-4200-1947}\and
           L.~Ferro~\inst{1,2}\orcidlink{0009-0006-1140-6913}\and
           F.~Frontera~\inst{1,2}\orcidlink{0000-0003-2284-571X}\and
           A.~Tsvetkova~\inst{5,2,6}\orcidlink{0000-0003-0292-6221}
          }
          
   \institute{
   Department of Physics and Earth Science, University of Ferrara, via Saragat 1, I--44122, Ferrara, Italy\label{unife}\and 
   INAF -- Osservatorio di Astrofisica e Scienza dello Spazio di Bologna, Via Piero Gobetti 101, I-40129 Bologna, Italy\label{oabo}\and 
   INFN -- Sezione di Ferrara, via Saragat 1, I--44122, Ferrara, Italy\label{infnfe}\and
   INAF, Osservatorio Astronomico d’Abruzzo, via Mentore Maggini snc, 64100 Teramo, Italy\and
   Department of Physics, University of Cagliari, SP Monserrato-Sestu, km 0.7, 09042 Monserrato, Italy\and
   Ioffe Institute, Politekhnicheskaya 26, 194021 St. Petersburg, Russia\and
   Astronomical Observatory of the Autonomous Region of the Aosta Valley (OAVdA), Loc. Lignan 39, I-11020, Nus (Aosta Valley), Italy\\
   }

  \date{Received 9 May 2024 / Accepted 5 July 2024}
  \abstract
  {The dissipation process responsible for the long gamma-ray burst (GRB) prompt emission and the kind of dynamics that drives the release of energy as a function of time are still key open issues. We recently found that the distribution of the number of peaks per GRB is described by a mixture of two exponentials, suggesting the existence of two behaviours that  turn up as peak-rich and peak-poor time profiles.}
  {Our aims are to study the distribution of the number of peaks per GRB of the entire catalogue of about 3000 GRBs observed by the {\it Fermi} Gamma-ray Burst Monitor (GBM) and to make a comparison with previous results obtained from other catalogues.}
  {We identified GRB peaks using the {\sc mepsa} code and modelled the resulting distribution following the same procedure that was adopted in the previous analogous investigation.}
  {We confirm that only a mixture of two exponentials can model the distribution satisfactorily, with model parameters that fully agree with those found from previous analyses. In particular, we confirm that $(21\pm4)$\% of the observed GRBs are peak-rich ($8\pm 1$ peaks per GRB on average), while the remaining 80\% are peak-poor ($2.12\pm0.10$ peaks per GRB on average).}
  {We confirm the existence of two different components, peak-poor and peak-rich GRBs, that make up the observed GRB populations. Together with previous analogous results from other GRB catalogues, these results provide compelling evidence that GRB prompt emission is governed by two distinct regimes. } 
   \maketitle

\section{Introduction}
\label{sec:intro}
Long-lasting gamma-ray bursts (GRBs) are thought to be produced by the collapse of the core of some hydrogen-stripped massive stars, in which a relativistic jet is launched by the newborn compact object, either a neutron star (NS) or a black hole (BH), and makes it to the stellar photosphere \citep{Woosley93,Paczynski98,MacFadyen99,Yoon05}. Today, almost 30 years since the first discoveries of the long-lasting multiwavelength counterpart (from gamma rays, X-rays, and optical light to radio waves) following the initial gamma-ray emission (so-called afterglow), the wealth of knowledge of the GRB phenomenon, such as the energetics, the jet structure, its Lorentz factor, the density of surrounding environment, and the properties of the progenitor stars has grown remarkably (see e.g. \citealt{Vaneerten18,SalafiaGhirlanda22,Levan16b} for reviews). However, the kind of mechanism and energy reservoir powering the GRB prompt emission, the nature of the dissipation process into gamma-rays, and the distance from the progenitor star at which it takes place, remain highly debated and hot topics (see \citealt{Peer15rev,Zhang18_book} for reviews).

Even inexperienced eyes are caught by the variety of prompt emission light curves (LCs), which range from simple and smooth single pulses all the way up to very complex spiky profiles including many peaks and occasionally quiescent times in between, during which the signal can temporarily drop below instrumental sensitivity. Ever since, attempts to interpret the variability and apparent lack of systematic temporal evolution of the most variable LCs have resulted in mechanisms based on internal dissipation of some kind of energy (either kinetic or magnetic) into gamma-rays through shocks within a wind of relativistic shells (see e.g. \citealt{KumarZhang15rev} for a review). 
However, the rich wealth of information hidden in the observed complexity of GRB LCs remains mostly unintelligible. In particular, an open question is whether this variety could be the result of a common stochastic process.
If positive, a precise characterisation of this process would reveal how GRB inner engines work, shedding light on the nature of the powering mechanism and on the nature of the engines themselves.
Under this assumption, a way to gain clues is by studying the distributions of a number of properties, such as the duration, energy, and luminosity of individual peaks that make up GRB LCs, as well as their waiting times. Possible evidence for self-organised criticality was found and discussed in analogy with other astrophysical transient sources, such as solar flares or magnetar bursts, which might also be ruled by similar kinds of instabilities \citep{WangDai13,Lyu20,Li23,Li23b,Maccary24}.

The distribution of the number of peaks within long GRB LCs has been studied and modelled in detail for the first time only recently \citep[hereafter G24]{Guidorzi24}. This distribution was studied in four different experiments: the Burst And Transient Source Experiment (BATSE; \citealt{Paciesas99}), on board the Compton Gamma-Ray Observatory; the Burst Alert Telescope (BAT; \citealt{Barthelmy05}) on board the {\it Neil Gehrels Swift Observatory}; the {\it BeppoSAX} Gamma-Ray Burst Monitor (GRBM; \citealt{Frontera97}); and the {\it Insight-HXMT} High Energy instrument (HE; \citealt{Liu20_HXMT}). 
The  result, which emerged from the analysis of these four independent catalogues, is that the distribution of the number of peaks is modelled by a mixture of two exponentials (M2E). The two components are characterised by two different average numbers of peaks per GRB: $2.1\pm 0.1$ and $8.3\pm 1.0$, nicknamed `peak-poor' and `peak-rich', respectively.
The result was unexpected, given also the diversity of the four experiments in terms of energy passbands and effective areas. The existence of two distinct components was interpreted as evidence of two correspondingly different dynamical regimes, through which long GRB inner engines work.

In their work, \citetalias{Guidorzi24} could not extend the same analysis to the entire GRB catalogue of the {\it Fermi} Gamma-ray Burst Monitor (GBM; \citealt{Meegan09}), but reported an analogous preliminary result obtained over a sample of nearly 400 GRBs.
In the present work we report the same kind of analysis applied to the full GRB catalogue of GBM currently available. The importance of this additional investigation cannot be overstated: not only does it represent a further test on independent data, but it is also statistically very sensitive, given that the GBM data set outnumbers the data sets previously analysed by~\citetalias{Guidorzi24}. In addition, since GBM and BAT share the largest number of GRBs, it is possible to further explore to what extent the number of peaks of any given GRB depends on the detector.

The paper is organised as follows. Section~\ref{sec:data} describes the data sample and reduction. Section~\ref{sec:res} presents the results of the statistical analysis, whose discussion and conclusions are reported in Section~\ref{sec:disc}.

\section{Data analysis}
\label{sec:data}

\subsection{Data set}
\label{sec:dataset}

The GBM is composed of $12$ NaI scintillators sensitive in the $8$--$1000$ keV energy range and of two BGO scintillators, sensitive to the high energy tail of a typical GRB spectrum ($150$ keV--$30$ MeV), overlapping with NaI detectors at low energies.
We started from $3091$ long 
GRBs detected by GBM from 14 July 2008 to 4 February 2024. We excluded short GRBs by discarding those with $T_{\rm{90}}<2$~s, where $T_{\rm{90}}$ is the time interval encompassing from 5\% to 95\% of the total fluence, as well as those with clear evidence of a compact merger origin, such as 211121A and 230307A\footnote{191019A should also appear in this list, but it was not seen by the GBM.} \citep{Levan23,Troja22,Gompertz23,Levan24}. Very bright GRBs that saturated the GBM detectors, such as 130427A and 221009A \citep{Preece14,Burns23}, were also left out. From the remaining GRBs, $154$ have a known (spectroscopic) redshift, and $410$ are in common with BAT.

Gamma-ray bursts affected by the simultaneous occurrence of a solar flare were discarded. Solar flares were identified by observing the relative intensity between the flare and the GRB on different units and on different energy ranges, taking into account the direction of the Sun. Being spectrally softer than GRBs, most solar flares are barely visible above 40~keV.

We finally rejected all the GRBs whose LC was lacking data points within the $T_{90}$ interval. Eventually, we ended up with 2971 GRBs, which is hereafter referred to as the GRB sample used for the analysis.

\subsection{Background subtraction}
\label{sec:bkg}
Gamma-ray burst LCs were extracted in the full NaI energy range ($8-1000$ keV), using a $64$ ms bin time.
We subtracted the background using the GBM data tools\footnote{\url{https://fermi.gsfc.nasa.gov/ssc/data/analysis/gbm/gbm_data_tools/gdt-docs/}.}

\citep{GbmDataTools} following standard prescriptions: we selected one time window preceding the burst and another one following, both having comparable or longer duration than that of the interval containing the burst,\footnote{This requirement ensures that all the harmonics that significantly contribute to the background in the burst interval can be adequately modelled in the selected adjacent windows.} and interpolated the background with a polynomial function of order up to 3. For each GRB we looked into the `scat detector mask' entry on the HEASARC catalogue\footnote{\url{https://heasarc.gsfc.nasa.gov/db-perl/W3Browse/w3table.pl?tablehead=name\%3Dfermigbrst&Action=More+Options}}
to identify the detectors used by the GBM team. We used the TTE (Time Tagged Event) data whenever they covered the whole GRB, from the start of its $T_{90}$ interval to the end.

We removed the spikes caused by charged particles hitting the detectors as follows: all the bins whose counts exceeded the neighbouring counts by $\ge 9\sigma$ were tagged as spike candidates. Once visual inspection of different GBM units confirmed the particle, and thus the spurious nature of a given spike candidate, its counts were replaced with the mean of the neighbouring bins.

The quality of the background subtraction was assessed by computing mean and standard deviation of the normalised residuals\footnote{They are defined as the difference between counts and model, divided by the corresponding uncertainty (see \citealt{Maccary24}).} of the time windows used for background interpolation. In particular, we made sure that the background-subtracted counts satisfied the uncorrelated Gaussian noise assumption required by {\sc mepsa} (\citealt{Guidorzi15a}; see Section~\ref{sec:peaks}) first by carrying out a normality test on the normalised residuals (Kolmogorov-Smirnov), to ensure that the noise was Gaussian, and then by examining the autocorrelation function (ACF) of the normalised residuals and by carrying out a runs test, to ensure they were uncorrelated.

\subsection{Identification of peaks}
\label{sec:peaks}
In line with~\citetalias{Guidorzi24}, all the 64 ms background-subtracted LCs were sifted with {\sc mepsa}, a well-calibrated code tailored to identify statistically significant GRB peaks in uniformly sampled time series affected by Gaussian uncorrelated noise. We discarded peaks with signal-to-noise ratio S/N$<5$. We ended up with 9625 peaks from 2954 GRBs having at least one significant peak. In parallel, we repeated the same selection by imposing two further thresholds on S/N: 8822 peaks from 2790 GRBs for S/N$>7$; 7371 from 2388 GRBs for S/N$>9$.

\section{Results}
\label{sec:res}
\subsection{Fermi-GBM sample}
\label{sec:GBM_sample}
\begin{figure}[h]
    \centering
    \includegraphics[width=0.45\textwidth]{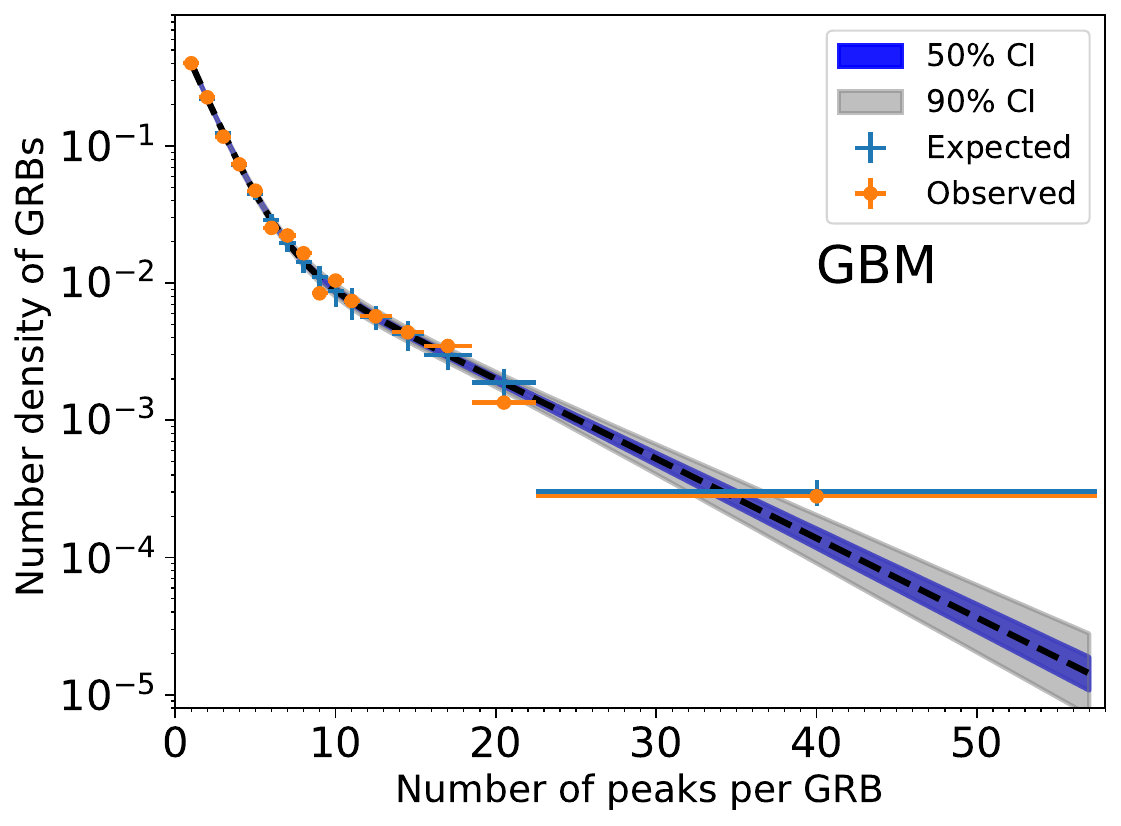}
    \caption{Distribution of the number of peaks per GRB for the \textit{Fermi/GBM} catalogue. The observed and expected counts are displayed in orange and blue, respectively. The histogram bins were grouped to ensure a minimum number of 15 expected counts per bin. The dashed black line shows the best-fitting model of mixture of two exponentials. The blue and grey regions respectively show the 50\% and 90\% model confidence interval obtained from sampling the posterior distribution of the model parameters computed through Markov chain Monte Carlo simulations.}
    \label{fig:npeak}
\end{figure}

We computed the distribution of the number of peaks per GRB, following the procedure by~\citetalias{Guidorzi24}, and tested the various models considered therein (simple exponential, simple and broken power law, and stretched exponential). As did \citetalias{Guidorzi24}, we found that the data can only be satisfactorily fitted by a M2E model given by \begin{equation}
    {f(n)\ =\ k\ ({e^{-n/n_1} \ +\ \xi \,e^{-n/n_2}})\;},
    \label{eq:exp2}
\end{equation}
where $k$ is a normalisation constant, $n_i$ is the characteristic number of peaks per GRB of the $i$-th
component ($i=1,2$), and $\xi$ is a relative normalisation parameter.
The expected number of peaks per GRB of the $i$-th component is given by
\begin{equation}
\langle n^{(i)} \rangle\ =\ \frac{\sum_{n=1}^{+\infty} n e^{-n/n_i}}{\sum_{n=1}^{+\infty} e^{-n/n_i}}\ = \frac{1}{1-e^{-1/n_i}}\;,
\label{eq:mean_exp}
\end{equation}
and the fraction of GRBs contributing to the second component is given by
\begin{equation}
    \bar{w}_2\ =\ k\ \xi \sum_{n=1}^{+\infty} e^{-n/n_2}\;.
    \label{eq:waver}
\end{equation} The observed distribution along with the best fit model are shown in Figure~\ref{fig:npeak}, while Table~\ref{Tab:results_GBM} reports the best fit values of the model parameters and the corresponding $p$-value of the $\chi^{2}$ test.

The M2E model fits the data very well (e.g. $p$-value of the two-tail $\chi^{2}$ test of $0.80$), while other models were unable to describe it (e.g. a fit with a stretched exponential yielded a $p$-value of 0.0026), thus confirming the previous results. The mean number of peaks for the peak-poor and for the peak-rich component is $\langle n^{(1)}\rangle = 2.10^{+0.10}_{-0.10}$ and $\langle n^{(2)}\rangle = 7.61^{+0.97}_{-0.84}$, respectively.
The fraction of peak-rich GRBs is $\bar{w}_2 = 0.21^{+0.04}_{-0.04}$.

\begin{table*}
\centering 
\begin{tabular}{lrrcccccccc}
\hline\hline
S/N  & $N$ & $N_{p}$ & $k$ & $n_1$ & $\langle n^{(1)}\rangle$ & $n_2$  & $\langle n^{(2)}\rangle$ & $\xi$ & $\bar{w}_2$ & $\chi^2$ test\\
Threshold & & & & & & & & & & $p$-value\\ 
\hline
$5$  &  $2954$ & $9625$ & $0.71^{+0.05}_{-0.05}$ & $1.55^{+0.11}_{-0.11}$ & $2.10^{+0.11}_{-0.11}$ & $7.07^{+0.99}_{-0.82}$ & $7.61^{+0.97}_{-0.84}$  & $0.04^{+0.01}_{-0.01}$ &$0.21^{+0.04}_{-0.04}$ & $0.80$ \\ 
$7$ & $2790$ & $8822$ & $0.82^{+0.06}_{-0.06}$ & $1.39^{+0.11}_{-0.10}$ & $1.95^{+0.10}_{-0.10}$ & $6.78^{+0.93}_{-0.77}$ & $7.53^{+0.90}_{-0.77}$  & $0.04^{+0.01}_{-0.01}$ &$0.23^{+0.04}_{-0.04}$ & $0.78$ \\ 
$9$ & $2388$ & $7371$ & $0.93^{+0.08}_{-0.08}$ & $1.26^{+0.10}_{-0.10}$ & $1.83^{+0.10}_{-0.10}$ & $6.70^{+0.93}_{-0.75}$ & $7.23^{+0.90}_{-0.77}$  & $0.039^{+0.012}_{-0.010}$ &$0.23^{+0.04}_{-0.04}$ & $0.43$ \\ 
\hline
\end{tabular}
\caption{Best-fit values and 90\% confidence intervals of the parameters of the model of the mixture of two exponentials applied to the \textit{Fermi/GBM} data set. All these parameters are defined in Section \ref{sec:res}.}
\label{Tab:results_GBM}
\end{table*}

\subsection{Swift-BAT/Fermi-GBM common sample}
\label{sec:common}
{\it Fermi}/GBM shares 410 GRBs with {\it Swift}/BAT, 361 of which were covered by BAT in burst mode.
As~\citetalias{Guidorzi24} did for the common sample between {\it CGRO}/BATSE and {\it BeppoSAX}/GRBM, here we used the BAT-GBM common sample to study how the number of peaks of any given GRB depends on the detector;  the ultimate goal was to test the robustness of our result.

\begin{figure}[htpb]
    \centering
\includegraphics[width=0.5\textwidth]{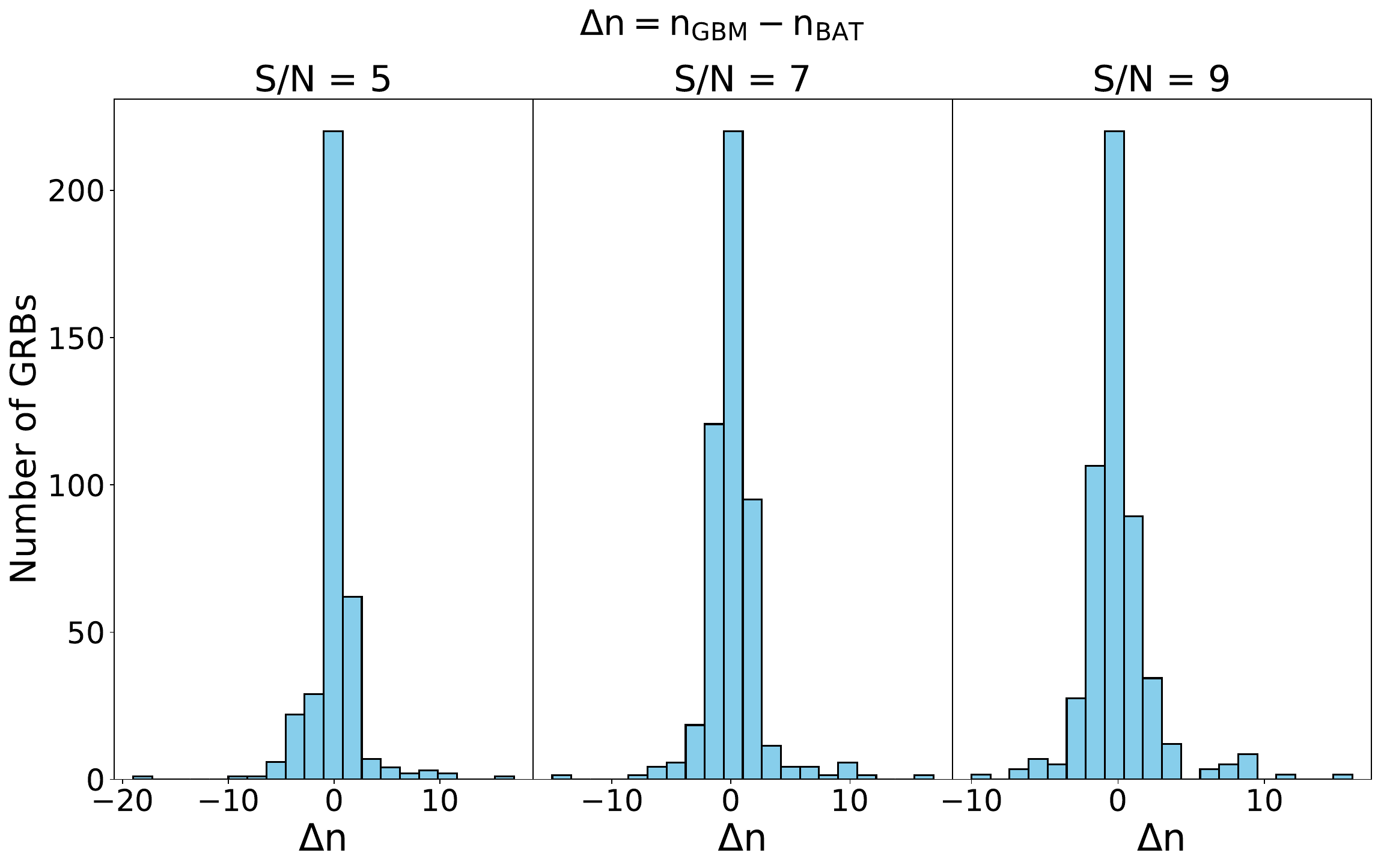}
    \caption{Distribution of $\Delta n$ for three different thresholds on S/N: 5, 7, and 9. }
    \label{fig:BAT_GBM_hist_deltan}
\end{figure}

Following \citetalias{Guidorzi24}, we considered the difference $\Delta n = n_{\rm{GBM}}- n_{\rm{BAT}}$
between the number of peaks detected by GBM and that of BAT for each common GRB. We replicated the same analysis assuming the three different thresholds on S/N: S/N $>5$; S/N$>7$; and S/N$> 9$. Figure~\ref{fig:BAT_GBM_hist_deltan} shows the distributions of $\Delta n$ for the three different thresholds on S/N.
The median value of $\Delta n$ is zero for all S/N thresholds. The interval $|\Delta n|\le 2$ collects from 85 to 88\% of the complete common samples, whereas from 92 to 94\% common GRBs have $|\Delta n|\le 3$. These values prove that both experiments are equivalently sensitive to the number of peaks, in spite of their different energy passbands.
As noted by~\citetalias{Guidorzi24}, $\Delta n$ for most GRBs is significantly smaller than the difference between the mean values of peak-rich and peak-poor GRBs, $\langle n^{(2)}\rangle - \langle n^{(1)}\rangle = 5.5^{+0.9}_{-0.7}$ (90\% confidence), which 
makes the identification of the two families of long GRBs robust and essentially detector-independent.
\subsection{Combined sample}
\label{sec:merged}
We merged the results of the five catalogues
and treated them as if they were a single result. The GBM result reported in the present work, is clearly dominant as its 2954 GRBs outnumber those from the the other catalogues (which included 1457, 1277, 820, and 202 GRBs). We obtained an acceptable fit ($p$-value of 0.06) with the same but more accurate value for the fraction of peak-rich GRBs: $\bar{w}_2~(\rm{all}) = 0.20^{+0.02}_{-0.02}$, and similar values for the mean number of peaks for the peak-poor and peak-rich component $\langle n^{(1)}\rangle = 2.25^{+0.08}_{-0.07}$ and $\langle n^{(2)}\rangle = 9.08^{+0.72}_{-0.66}$, respectively.

\subsection{Impact of selection effects and the Malmquist bias}
\label{sec:sel_effects_bias}
 The number of peaks per GRB clearly depends on the S/N of each peak. As such, the selection of GRBs with suitable peaks as well as their classification of peak-poor versus peak-rich depends on S/N, too. 
Any given GRB that would have been observed at higher-than-actual redshift, could have been either tagged with fewer peaks or missed altogether. 
Consequently, our results are potentially affected by a significant Malmquist bias. As we also did in \citetalias{Guidorzi24}, repeating the analysis for a range of different S/N thresholds, and thus mimicking the effect of moving given GRBs further away, our aim was to explore the robustness of our results against any distance-related bias.
Two interesting facts emerge from this analysis:
\begin{enumerate}
    \item The fraction $\bar{w}_2$ of peak-rich GRBs essentially remains constant for all catalogues and for the different S/N thresholds, and therefore appears to be a robust property that is least affected by anything that depends on peak brightness. In the following, we delve further into this aspect, which seems to clash with the Malmquist bias;
    \item The average number of peaks per GRB changes only slightly for different thresholds on S/N: from $2.1$ (S/N>5) to $1.8$ (S/N>9) for the peak-poor class, and, correspondingly, from $7.6$ to $7.2$ for the peak-rich class (Table~\ref{Tab:results_GBM}). This is understood by the following argument: not only does the number of peaks decrease by increasing the threshold on S/N of any given GRB (mimicking the consequence of moving it to higher redshifts), but   the number of GRBs featuring at least one significant peak also decreases. As a result, the average number of peaks per surviving GRB decreases less abruptly than the overall number of peaks.
\end{enumerate}

In principle, a direct way to understand the impact of the Malmquist bias is by focusing on a sample of GRBs with measured redshift. Unfortunately, this sample is currently too small to make a statistically sound analysis feasible. Nonetheless, we explored in more detail the reasons for the first consideration above as follows. We addressed two questions: By moving a peak-rich GRB to progressively higher redshifts, we wanted to know (a) at what relative distance it becomes peak-poor, thus contributing to altering the true distribution, and (b) at what relative distance the GRB disappears, an event coinciding with the brightest peak sinking below threshold?

The answer to (b) depends on the intensity of the brightest peak, while we assume that the answer to (a) depends on the intensities of the third or fourth brightest peaks, once we take three or four as a reasonable proxy of the number of peaks for discriminating between peak-poor and peak-rich GRBs. 

For each of the peak-rich GRBs with at least 6 S/N$>5$ peaks, we took the four brightest ones. We ended up with 399 GRBs. For each GRB we calculated the three following ratios: $r_{21}\ge r_{31}\ge r_{41}$, respectively corresponding to the ratio of the intensities of the second, of the third, and of the fourth brightest peaks to the intensity of the brightest peak.


As a result, on average it is $r_{41}\sim 2/3$, with $0.56$--$0.77$ as interquartile, so the range of intensities of the four brightest peaks is relatively narrow.
Specifically, from the disappearing of the fourth peak to the disappearing of the brightest peak, the threshold on S/N increases only by $\sim$50\% on average. Translating the relative change of the S/N threshold in terms of the relative change in the maximum volume within which a given peak can be detected, the volume increases by a factor of $(2/3)^{-3/2}\approx 1.8$, so 80\% (we used the scaling $V\propto p^{-3/2}$ of volume $V$ on peak flux $p$).
This is a relatively short range compared with the range of redshifts spanned by the GRB populations that make up the observed catalogues and the GBM catalogue in particular. Replacing the fourth with the third brightest peak, the relative volume change shrinks to 60\%, so the previous conclusion holds true a fortiori. This property of most peak-rich GRBs may explain why the Malmquist bias does not strongly affect the result of a constant peak-rich fraction throughout the different catalogues and GRB populations.

\section{Discussion and conclusions}
\label{sec:disc}
\citet{Guidorzi24} modelled for the first time the distribution of the number of peaks in long GRB LCs through the analysis of four independent catalogues ({\it CGRO}/BATSE, {\it Swift}/BAT, {\it BeppoSAX}/GRBM, and {\it Insight-HXMT}/HE), which differ from each other in terms of energy passbands and effective areas. Only a mixture of two exponentials was found to model all four distributions, supporting the existence of two families, called peak-poor and peak-rich, characterised by an average number of peaks per GRB of around $2.1\pm 0.1$ and $8.3\pm 1.0$, respectively. It was also found that peak-poor GRBs make up about 80\% of the GRB populations observed by past and present experiments.

In the present work we carried out the analogous investigation over the {\it Fermi}/GBM GRB catalogue, which currently includes about 3000 GRBs and therefore offers a further opportunity to test the previous results with a high degree of statistical accuracy.
Remarkably, not only did we confirm that the distribution can only be modelled by a mixture of two exponentials, but we also found that the best-fitting parameter values are fully consistent with what was found by~\citetalias{Guidorzi24} from different and independent data sets.
Specifically, the expected numbers of peaks for the two families of peak-poor and of peak-rich GRBs are remarkably similar to the~\citetalias{Guidorzi24} results. Analogously, the fraction of peak-rich GRBs is found to be $0.21\pm 0.04$, fully consistent with what was found by~\citetalias{Guidorzi24}. Although they were ignored in the present analysis, long-merger candidates would belong to the hard tail of the peak-rich family: 211211A has 48 peaks and 230307A has 53 peaks. Should peak richness be confirmed in future analogous events, it would provide a clue to the way the inner engines of these mergers work.

\begin{figure*}[h]
    \centering
    \includegraphics[width=0.9\textwidth]{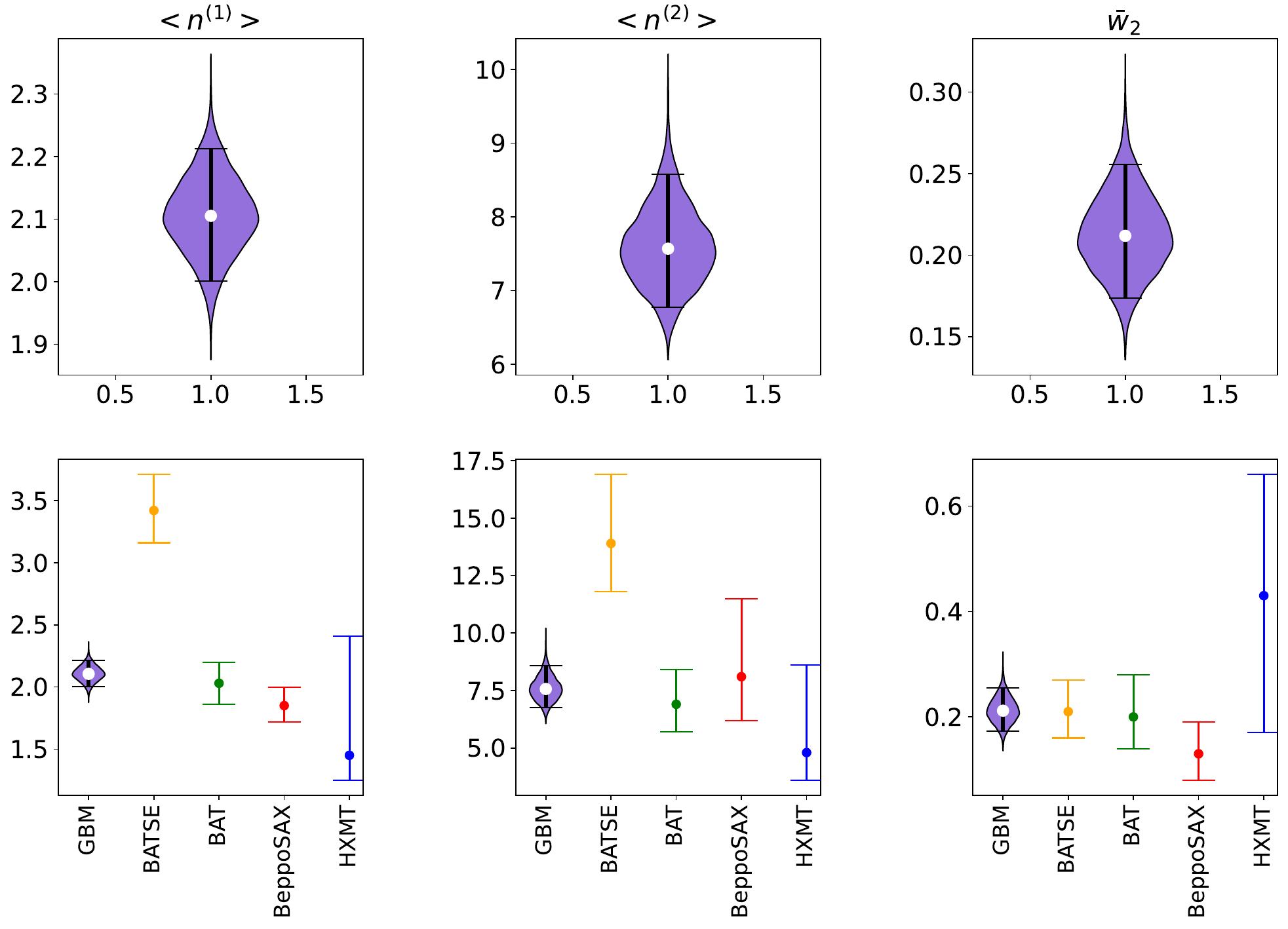}
    \caption{This figure describes our present results compared with those of \citetalias{Guidorzi24}. \textit{Top panels, left to right}: Violin plots of the posterior distributions of the three parameters $\langle n^{(1)}\rangle$, $\langle n^{(2)}\rangle$, and $\bar{w}_2$, respectively representing the expected number of peaks per GRB of the peak-poor and the peak-rich classes, and the fraction of peak-rich GRBs obtained for the GBM sample. The horizontal bars span 5 to 95\% quantiles. \textit{Bottom panels, left to right:} Comparison between the results obtained for the same parameters in this work, with those obtained by~\citetalias{Guidorzi24} from other catalogues.}
    \label{fig:violin_plot}
\end{figure*}

Figure~\ref{fig:violin_plot} shows a visual comparison between the GBM results reported here and the results obtained by~\citetalias{Guidorzi24}.
Interestingly, while the fraction of peak-poor GRBs is basically the same for all the five catalogues, the average numbers of peaks per each GRB for both families of the only BATSE is somewhat higher. As discussed in~\citetalias{Guidorzi24}, this is ascribed to the larger effective area of BATSE compared with the other instruments, including GBM.

In conclusion, we confirmed that the existence of these two families is a robust result that is not sensitive to the experiments and their energy passbands. As discussed by~\citetalias{Guidorzi24}, these two different dynamics ruling long GRB prompt emission might point to either two different families of engines left over by the collapse of the core, either a NS or a BH, or two different regimes in which GRB engines release their energy. As noted in~\citetalias{Guidorzi24}, the distinctive property of peak-rich GRBs is the presence of sub-second variability that adds to the slow-varying component, which is instead observed in both kinds of GRBs.

The existence of two distinct dynamics in the observed population of GRBs can be interpreted within different models of how dissipation into gamma-rays takes place (see~\citetalias{Guidorzi24} for a detailed discussion). Consequently, it does not directly help to discriminate between different interpretations. Nevertheless, it provides a solid clue to how GRB engines work, with the possibility that they could operate close to a critical regime. In this way, a completely different dynamical behaviour can result from a small difference in some key properties, such as the degree of magnetisation of the relativistic ejecta, as discussed in~\citetalias{Guidorzi24}.
Should the peak richness of GRBs be found preferentially associated with other key properties in future investigations, the implications would help to narrow down the identity of GRB engines and the nature of the dissipation process that governs the prompt emission.

\begin{acknowledgements}
We thank the anonymous reviewer for providing insightful comments that improved the manuscript.
 R.M. and M.M. acknowledge the University of Ferrara for the financial support of their PhD scholarships. R.M. acknowledges the Fermi team for covering flight costs for attending the Fermi summer school 2023.
 A.T. acknowledges financial support from ASI-INAF Accordo Attuativo HERMES Pathfinder operazioni n. 2022-25-HH.0 and the basic funding program of the Ioffe Institute FFUG-2024-0002. L.A. acknowledges support from INAF Mini-grant programme 2022. A.C. is partially supported by the 2023/24 ``Research and Education" grant from Fondazione CRT. The OAVdA is managed by the Fondazione Clément Fillietroz-ONLUS, which is supported by the Regional Governement of the Aosta Valley, the Town Municipality of Nus and the "Unité des Communes valdotaines Mont-Emilius".

\end{acknowledgements}


%
%
\bibliographystyle{aa}
\bibliography{bibliography,software,alles_grbs}
\end{document}